\documentclass{aastex631}

\begin{document}

\title{The First In-depth Photometric Study of the Four $\delta$ Scuti Stars Using TESS Data}

\author{Atila Poro}
\altaffiliation{Corresponding author: atilaporo@bsnp.info}
\affiliation{LUX, Observatoire de Paris, CNRS, PSL, 61 Avenue de l'Observatoire, 75014 Paris, France}
\affiliation{Astronomy Department of the Raderon AI Lab., BC., Burnaby, Canada}

\author{Hossein Azarara}
\affiliation{Physics Department, Shahid Bahonar University of Kerman, Kerman, Iran}

\author{Ahmad Sarostad}
\affiliation{Yazd Desert Night Sky Astronomy Institute, Yazd, Iran}

\author{Nazanin Kahali Poor}
\affiliation{Independent Astrophysics Researcher, Tehran, Iran}

\author{Razieh Aliakbari}
\affiliation{Physics Society of Iran (PSI), Tehran, Iran}

\author{Sadegh Nasirian}
\affiliation{Thaqib Astronomical Association, Rasht, Iran}

\author{Sahar Momeni}
\affiliation{Independent Astrophysics Researcher, Tehran, Iran}

\begin{abstract}
The first in-depth photometric study of four $\delta$ Scuti stars was performed. We used time series data from the Transiting Exoplanet Survey Satellite (TESS) that is available in different sectors. According to the extracted maxima from TESS space-based observations, we calculated an ephemeris for each star. We estimated the physical parameters of the target stars based on the Gaia Data Release 3 (DR3) parallax method. The results obtained for the surface gravity of the stars are consistent with the reports of the TESS Input Catalog and Gaia DR3. We estimated the pulsating constant based on the physical parameters and period of the stars. Therefore, we found that the stars 2MASS 15515693-7759002 and 2MASS 07513202+0526526 belong to the fundamental, while 2MASS 00044615+4936439 and 2MASS 10215638-3326137 relate to the first overtone. The Fourier analysis using the Period04 program was done for each star. As we showed in the Hertzsprung-Russell (H-R) diagram, the stars are located in the instability strip of the $\delta$ Scuti stars region. Four target stars were found to be of the low-amplitude $\delta$ Scuti star type.
\end{abstract}


%%%%%%%%%%%%%%%%% BODY OF PAPER %%%%%%%%%%%%%%%%%%
\section{Introduction}
The $\delta$ Scuti stars are a type of variable star characterized by pulsation periods ranging from 0.02 to 0.25 days. These stars are well-known for their intermediate masses and their complex pulsating behavior. $\delta$ Scuti stars are primarily found in the lower region of the Cepheid instability strip, which are mostly located on the main sequence of stellar evolution. The spectral types of $\delta$ Scuti stars range from A to F. They have effective temperatures ranging from approximately 6300 K to 8600 K. They can exist in various evolutionary stages, with masses between approximately 1.6 $M_\odot$ for shorter-period stars and about 2.4 $M_\odot$ for longer-period stars (\citealt{2011AJ....142..110M}[1]). Most $\delta$ Scuti stars exhibit radial and/or non-radial low-order pressure ($p$) mode pulsations. These variable stars can be classified into two main groups: High Amplitude $\delta$ Scuti Stars (HADS) and Low Amplitude $\delta$ Scuti Stars (LADS). HADS are characterized by variations in brightness with amplitudes greater than 0.3 magnitudes, while LADS exhibit brightness variation amplitudes of less than 0.1 magnitudes in the $V$ band (\citealt{1996AA...307..539R}[2], \citealt{2017NewA...54...86J}[3]). HADS are typically sub-giant stars, while LADS are found on or near the main sequence in the H-R diagram. Variable stars such as SX Phoenicis (SX Phe) type stars are also characterized by rapid and low-amplitude variations in brightness. They are considered to be older age stars than $\delta$ Scuti stars. These stars are located on the instability strip of the H-R diagram and are found in globular clusters as well as in old populations of the galactic field. Other types of pulsators that have low-amplitude pulsations are $\gamma$ Doradus ($\gamma$ Dor). These stars usually have longer periods than $\delta$ Scuti stars. $\gamma$ Dor stars are mostly F spectral type, so they are usually found in a region slightly cooler than $\delta$ Scuti stars. They are situated on or near the intersection of the classical instability strip and the main sequence on the H-R diagram.

In this investigation, we presented a photometric analysis for four $\delta$ Scuti stars. These stars were chosen randomly from the available short-period pulsation stars, considering they are $\delta$ Scuti star candidates, even though it is highly probable in some catalogs. The target stars are low-amplitude $\delta$ Scuti type star candidates that are worthy of attention. Furthermore, the exact type of some target stars is not precisely specified. We utilized the high-precision photometric observations from the TESS space telescopes to study $\delta$ Scuti stars, which exhibit subtle brightness changes. We determined the ephemeris of the stars by calculating the periods of star oscillations and analyzing the structure of the light curves and pulsation behavior. The physical and fundamental parameters were also obtained.

%%%%%%%%%%%%%%%%%%%%%%%%%%%%%%%%%%%%%%%%%%%%%%%%%%
\vspace{0.6cm}
\section{Target Stars and Dataset}
The target stars in this study are 2MASS 00044615+4936439, 2MASS 10215638-3326137, 2MASS 15515693-7759002, and 2MASS 07513202+0526526. They are classified in the ASAS-SN catalog as $\delta$ Scuti star. However, the ASAS-SN catalog places them in this classification with about a $99.3\%$ probability. Although the target stars are considered potential candidates with a high probability in the ASAS-SN catalog, more investigation is required to be confirmed.
2MASS 10215638-3326137, 2MASS 15515693-7759002, and 2MASS 07513202+0526526 are classified as $\delta$ Scuti stars in the AAVSO International Variable Star Index (VSX) database, ASAS-SN catalog, and the \cite{2022yCat.1358....0G}[4] study. Their period reported by ASAS-SN as 0.085428 day, 0.068385 day, and 0.064306 day for 2MASS 10215638-3326137, 2MASS 15515693-7759002, and 2MASS 07513202+0526526, respectively.
2MASS 00044615+4936439 has been identified as a $\delta$ Scuti star in most sources, similar to other target stars, but is solely categorized in the SuperWASP variable stars catalog (\citealt{2021MNRAS.502.1299T}[5]) as a contact binary star system. SuperWASP variable stars catalog reported 0.24 day for its orbital period, while ASAS-SN presented 0.11769 day for this target. 2MASS 00044615+4936439 is also classified as $\delta$ Scuti, $\gamma$ Dor, and SX Phe type stars in the online data catalog of the Gaia DR3 Part 4 Variability (\citealt{2022yCat.1358....0G}[4]), which indicates that the exact type of this star is uncertain in this catalog. The characteristics of the target stars from the Gaia DR3 and TESS sources are presented in Tables \ref{starinfoGaia} and \ref{starinfoTESS}, respectively.

Time series and high-quality TESS data were available for the target stars. NASA launched TESS with the goal of discovering exoplanets (\citealt{2010AAS...21545006R}[6], \citealt{2015JATIS...1a4003R}[7]). However, this telescope also provided researchers with high-precision photometric data from $\delta$ Scuti stars (\citealt{2018AJ....156..102S}[8]). TESS is equipped with four wide-field cameras that enable it to observe various regions of the sky. Each sector is observed for 27.4 days, allowing for an in-depth examination of specific areas. TESS presented data in the Barycentric Julian Date in Barycentric Dynamical Time ($BJD_{TDB}$).
TESS time-series are accessible through the Mikulski Archive for Space Telescopes (MAST)\footnote{\url{https://mast.stsci.edu/portal/Mashup/Clients/Mast/Portal.htmL}}. We extracted TESS-style light curves from MAST using the LightKurve\footnote{\url{https://lightkurve.github.io/lightkurve/}} software. The data was detrended using the TESS Science Processing Operations Center (SPOC) pipeline (\citealt{2016SPIE.9913E..3EJ}[9]). Table \ref{starinfoTESS} lists the TESS sectors used in this study.

\begin{table*}
\caption{Specifications of the target stars from the Gaia DR3.}
\centering
\begin{center}
\footnotesize
\begin{tabular}{c c c c c c}
\hline
Star & RA$.^\circ$(J2000) & Dec$.^\circ$(J2000) & $d$(pc) & $T$(K) & $log(g)$(cgs)\\
\hline
2MASS 00044615+4936439 & 1.192329 & 49.612196 & 1304(47) & 6705(47) & 3.558(1)\\
2MASS 10215638-3326137 & 155.484877 & -33.437133 & 1581(35) & 7066(13) & 3.685(11)\\
2MASS 15515693-7759002 & 237.987296 & -77.983485 & 1339(18) & 7199(22) & 3.992(17)\\
2MASS 07513202+0526526 & 117.883443 & 5.447915 & 1261(44) & 7879(31) & 4.094(23)\\
\hline
\end{tabular}
\end{center}
\label{starinfoGaia}
\end{table*}

\begin{table*}
\caption{Specifications of the target stars from the TESS.}
\centering
\begin{center}
\footnotesize
\begin{tabular}{c c | c c c | c c c}
\hline
2MASS & TIC & $T$(K) & $log(g)$(cgs) & $V$(mag.) & Sector & Exposure length(s) & Observation year\\
\hline
00044615+4936439 & 201575267 & 6783(129) & 3.448(139) & 11.942(24) & 17-57 & 1800-200 & 2019-2022\\
10215638-3326137 & 71490869 & 7089(130) & 3.638(95) & 12.464(57) & 36-62-63 & 600-200-200 & 2021-2023-2023\\
15515693-7759002 & 407224779 & 7185(150) & 4.123(89) & 13.326(92) & 65-66 & 200-200 & 2023-2023\\
07513202+0526526 & 271497849 & 7744(130) & 4.126(100) & 12.618(69) & 61 & 200 & 2023\\
\hline
\end{tabular}
\end{center}
\label{starinfoTESS}
\end{table*}

%%%%%%%%%%%%%%%%%%%%%%%%%%%%%%%%%%%%%%%%%%%%%%%%%%
\vspace{0.6cm}
\section{Ephemeris Calculation}
Calculating the ephemerides of $\delta$ Scuti stars is a common method for determining period changes. It shows the changes in maximum times over time compared to the reference ephemeris. So, we calculated the period ($P_0$) of the stars using the Period04 program (\citealt{2004IAUS..224..786L}[10]). Period04 was designed for analyzing time series data and employs various algorithms to analyze the data, including Fourier analysis. We extracted times of maxima for all available time-series TESS data. We obtained times of maxima by fitting a Gaussian curve to each maximum in the light curve. An appropriate maximum time from the TESS data was selected as $t_0$. Then, we plotted a linear fit of the maxima times based on epochs and calculated the ephemeris for each star. We presented ephemeris for each target star in Table \ref{ephemeris}. It should be noted that the low-amplitude $\delta$ Scuti stars are not very stable pulsators, and linear ephemerides might not work for sufficiently long time intervals. Therefore, it is required to conduct more observations in the future for more investigations.

\begin{table*}
\caption{The calculated ephemeris for target stars. The maximum time is in $BJD_{TDB}$, and the period is in day units.}
\centering
\begin{center}
\footnotesize
\begin{tabular}{c c c}
\hline
Star && Ephemeris\\
\hline
2MASS 00044615+4936439 && $2459853.44723(30)+0.117673472(56) \times E$ \\
2MASS 10215638-3326137 && $2459281.90989(16)+0.085429586(21)\times E$ \\
2MASS 15515693-7759002 && $2460097.75051(11)+0.068384840(553) \times E$ \\
2MASS 07513202+0526526 && $2459963.83413(16)+0.064305814(737)\times E$ \\
\hline
\end{tabular}
\end{center}
\label{ephemeris}
\end{table*}

%%%%%%%%%%%%%%%%%%%%%%%%%%%%%%%%%%%%%%%%%%%%%%%%%%
\vspace{0.6cm}
\section{Physical Parameters}
We utilized the Gaia DR3 parallax method to estimate the absolute parameters of our target stars (\citealt{2024NewA..11002227P}[11]). This method is a suitable choice when photometric data is available, thanks to the high accuracy of parallax measurements provided by Gaia DR3. Estimating absolute parameters using Gaia DR3 parallax has limitations, and requires a low total extinction ($A_V$) value (\citealt{2024PASP..136b4201P}[12]). We calculated the $A_V$ value using the 3D dust-map Python package considering the Gaia DR3 reported distance (\citealt{2019ApJ...887...93G}[13]). The results of $A_V$ were in an acceptable state for using the Gaia DR3 parallax method to estimate the absolute parameters of stars.
First, the method estimated the absolute magnitude of the star ($M_V$) by using the apparent magnitude in the $V$ filter, the distance from Gaia DR3 ($d$) in parsec unit, and $A_V$. We used $V$ magnitude from the ASAS-SN database. Then, the bolometric correction ($BC$) from the \cite{1996ApJ...469..355F}[14] study, corrected by \cite{2010AJ....140.1158T}[15], was used to derive the bolometric absolute magnitude ($M_{bol}$) for each star. We calculated the luminosity ($L$) using Pogson's relation (\citealt{1856MNRAS..17...12P}[16]) and $M_{bol}$. Gaia DR3 provides spectroscopic and photometric data and measurements of a star's temperature (\citealt{2016AA...595A...1G}[17]). The effective temperature ($T$) reported by Gaia DR3 enables us to calculate the radius ($R$). We considered an average of the upper and lower uncertainties of the temperature reported by Gaia DR3.
We estimated the mass of stars ($M$) using the relationship presented in the \cite{cox2015allen}[18] study (Equation \ref{eqM}).

\begin{equation}\label{eqM}
logM=0.46-0.1M_{bol}
\end{equation}

We estimated the surface gravity ($g$), which was in good agreement with the values reported by TESS and Gaia DR3 for each star (Equation \ref{eqg}).

\begin{equation}\label{eqg}
g=G_{\odot}(M/R^2)
\end{equation}

\begin{table*}
\caption{The fundamental parameters of the four $\delta$ Scuti Stars.}
\centering
\begin{center}
\footnotesize
\begin{tabular}{c | c c c c c c | c c}
\hline
Star & $M_V$(mag.) & $M_{bol}$(mag.) & $L$($L_{\odot}$) & $R$($R_{\odot}$) & $M$($M_{\odot}$) & $log(g)$(cgs) & $A_V$ & $BC$\\
\hline
2MASS 00044615+4936439 & 1.208(55) & 1.228(55) & 25.174(1.241) & 3.726(38) & 2.174(28) & 3.633(3) & 0.158(2) & 0.020\\
2MASS 10215638-3326137 & 1.236(8) & 1.268(8) & 24.247(188) & 3.293(1) & 2.154(4) & 3.736(1) & 0.233(1) & 0.032\\
2MASS 15515693-7759002 & 2.474(62) & 2.508(62) & 7.740(430) & 1.792(38) & 1.619(23) & 4.152(12) & 0.218(1) & 0.034\\
2MASS 07513202+0526526 & 2.065(6) & 2.094(6) & 11.330(67) & 1.810(9) & 1.781(3) & 4.173(4) & 0.049(1) & 0.029\\
\hline
\end{tabular}
\end{center}
\label{physiclparameters}
\end{table*}

%%%%%%%%%%%%%%%%%%%%%%%%%%%%%%%%%%%%%%%%%%%%%%%%%%
\vspace{0.6cm}
\section{Results and Conclusion}
We conducted the first in-depth photometric study on four $\delta$ Scuti stars, including ephemeris calculation and estimating physical parameters. We used the TESS data for this investigation. Based on the results, the following are presented as discussions and conclusions:

A) We characterized the pulsations of the stars by means of a standard Fourier analysis of their light curves. We utilized Period04, a statistical analysis program that offers capabilities to extract the individual frequencies from the multi-periodic content of time series and provides a customization interface to do frequency fits (\citealt{2004IAUS..224..786L}[10]).

\begin{figure*}
    \centering
    \includegraphics[scale=0.2]{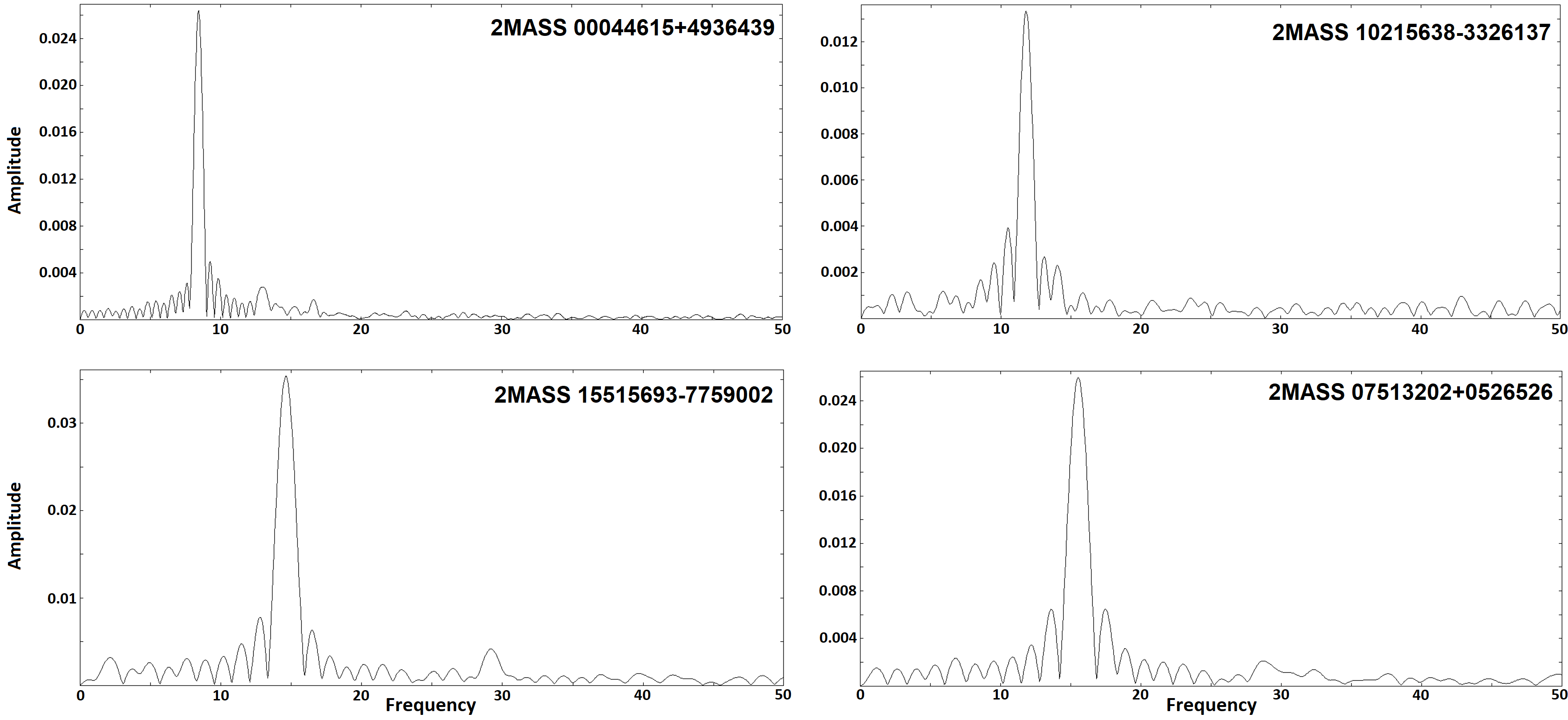}
    \caption{The amplitude-frequency diagram of the target stars.}
    \label{ferq}
\end{figure*}

\begin{figure*}
    \centering
    \includegraphics[scale=0.28]{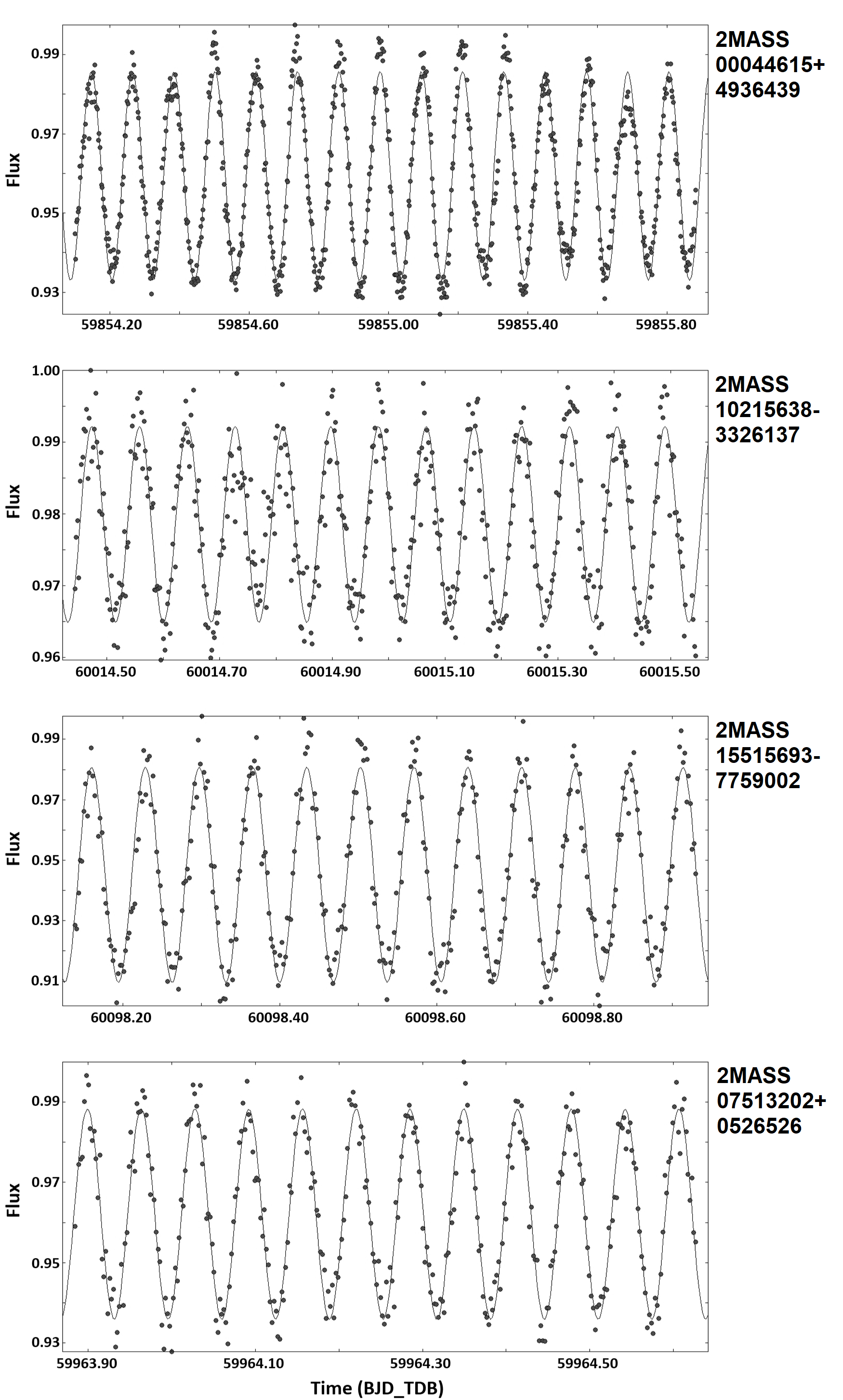}
    \caption{Fitted light curves using the main frequencies with the Period4 program. The time axis is subtracted from 2,400,000.}
    \label{LC-Fits}
\end{figure*}

The Fourier analysis of the stars are shown in Figure \ref{ferq}. Using the main oscillation mode, we fitted the light curve as in Figure \ref{LC-Fits} on the TESS data for each star. All frequencies and their amplitudes are provided in Table \ref{Q-F-Amp}.

\begin{figure*}
    \centering
    \includegraphics[scale=0.26]{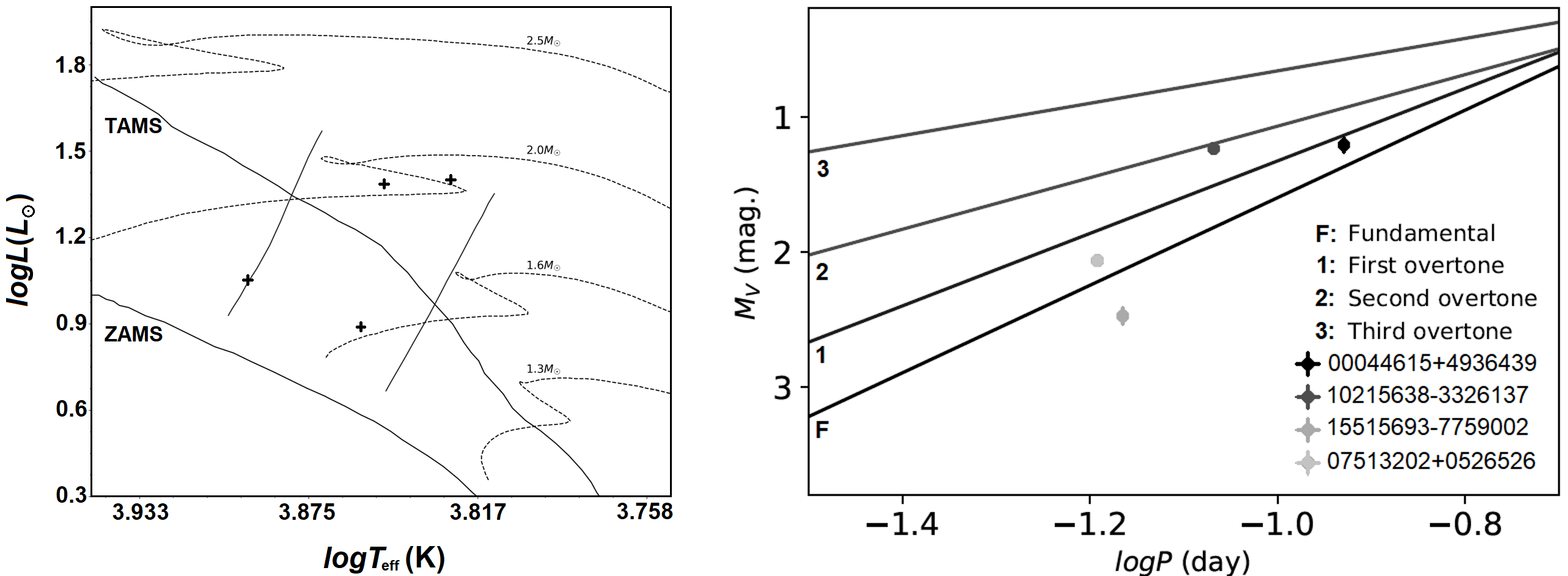}
    \caption{Left: Location of target stars on the H-R diagram. The instability strip of the $\delta$ Scuti stars is shown in black color lines. The evolutionary tracks were taken from the study of \cite{2016MNRAS.458.2307K}[31]. Right: $logP-M-V$ diagram and linear fits are from the \cite{2024RAA....24b5011P}[24] study. Star names in the figure legend begin with 2MASS.}
    \label{HR-PMv}
\end{figure*}

B) Using Gaia DR3, and TESS Input Catalog v8.2 database, we estimated the physical parameters of target stars. The results of the $log(g)$ parameter show good agreement with the TESS and Gaia DR3 findings. According to the findings, the positions of the stars in the H-R diagram are displayed in Figure \ref{HR-PMv}left. All four target stars are positioned between the two theoretical lines of the $\delta$ Scuti star range. 2MASS 00044615+4936439 had an uncertain exact type compared to the other targets in the catalogs, which is classified as a $\delta$ Scuti star in this study based on its location on the H-R diagram as well as other physical characteristics.

C) $\delta$ Scuti stars' pulsations can be explained through the pressure ($p$) and gravity ($g$) modes. $p$ modes usually have higher frequencies compared to $g$ modes. $p$ modes predominantly influence the outer atmosphere of the star, which leads to observable variations in brightness. $g$ modes are often trapped in the interior of the star, leading to oscillations that may not be as visible as the $p$ modes. $g$ modes typically exhibit more damped oscillations because they penetrate deeper into the star where energy is dissipated. Generally, in the $p$ mode, pressure acts as a restoring force by sound waves with vertical movements for a star that has lost its equilibrium condition, and in the $g$ mode, gravity functions as a restoring force with horizontal movements for the star (\citealt{2015JApA...36...33J}[19]). To study the interior structure of $\delta$ Scuti stars, identifying the $p$ mode order is required, and the pulsating constant ($Q$) is a crucial variable. $Q$ can be estimated by Equation \ref{eqQ} based on the \cite{1990DSSN....2...13B}[20] and \cite{2021PASP..133h4201P}[21] studies as

\begin{equation}\label{eqQ}
log(\frac{Q}{P})=0.1(M_{bol}-M_{bol\odot})+0.5log(\frac{g}{g_{\odot}})+log(\frac{T_{eff}}{T_{eff\odot}})
\end{equation}

\noindent where $P$ is the period of pulsation in days. As is clear from Equation \ref{eqQ}, the value of $Q$ depends on the fundamental parameters of the star. According to the \cite{1990DSSN....2...13B}[20] study, the measurement of $Q$ can have a fractional uncertainty of up to $18\%$. Equation \ref{eqQ} was used to calculate each star's $Q$ value (Table \ref{Q-F-Amp}). For this calculation, we considered the period from Table \ref{ephemeris}, parameters $M_bol$, $log(g)$ from Table \ref{physiclparameters}, and $T_{eff}$ from Table \ref{starinfoGaia}. We have determined the $p$ mode order for each pulsating star. Based on the value of $Q$ and consideration of $p$ mode, we can find that our stars are the fundamental or first-, second-, or third-overtones from the \cite{1979PASP...91....5B}[22], \cite{1997ESASP.402..367N}[23], and \cite{2024RAA....24b5011P}[24] studies (Table \ref{Q-F-Amp}). The following describes each group's range:

\begin{equation}\label{eq9}
\left\{\begin{array}{l}
Fundamental: 0.027 \leq Q\\
First\ overtone: 0.021 \leq Q < 0.027\\
Second\ overtone: 0.018 \leq Q < 0.021\\
Third\ overtone: Q < 0.018
\end{array}\right.
\end{equation}

As listed in Table \ref{Q-F-Amp}, target stars have the same results in all three sources mentioned.

D) One of the fundamental properties of pulsating variable stars is the period-absolute magnitude (generally called $P-L$) relationship that was first introduced by the \cite{1912HarCi.173....1L}[25] study. The first $P-L$ relationship for $\delta$ Scuti stars was established based on their adherence to the same $P-L$ relationship for Cepheid variables. To improve this $P-L$ relationship and utilize it as a standard candle, \cite{1992AJ....103.1647F}[26] investigated the combination of several Cepheids and $\delta$ Scuti stars. In the following decades, subsequent studies aimed to enhance the $P-L$ relationship for $\delta$ Scuti stars, such as \cite{2011AJ....142..110M}[1], \cite{2020MNRAS.493.4186J}[27], \cite{2021PASP..133h4201P}[21], \cite{2022MNRAS.516.2080B}[28], and \cite{2022ApJ...940L..25M}[29], \cite{2024RAA....24b5011P}[24].
The \cite{2024RAA....24b5011P}[24] study used least squares linear regression to update fundamental and three overtone modes relationships employing 2375 $\delta$ Scuti stars as a sample. The locations of four stars on the $P-M_V$ diagram are displayed in Figure \ref{HR-PMv}right. The two less massive stars are located near the fundamental linear fit, and the two more massive stars are located in the overtones range (Table \ref{physiclparameters}, Figure \ref{HR-PMv}right).

E) According to the brightness variation amplitudes of the target stars (Table \ref{Q-F-Amp}) as well as their position in the H-R diagram (Figure \ref{HR-PMv}left), they are of the LADS-subtype $\delta$ Scuti stars. LADS-subtype $\delta$ Scuti stars are found in or near the MS region in the H-R diagram. These subtype stars are usually of intermediate mass, typically ranging from about 1.5 to 2.5 solar masses, which is consistent with our results (Table \ref{physiclparameters}). The pulsations in LADS are driven by the $\kappa$ (kappa) mechanism, which involves the ionization of helium in the outer layers of the star; This process causes variations in opacity and leads to oscillations (\citealt{2022RAA....22j5006X}[30]).

\begin{table*}
\caption{Estimated $Q$ Value (left side), and obtained frequency and amplitude (right side) for the target stars.}
\centering
\begin{center}
\footnotesize
\begin{tabular}{c | c c c c | c c}
\hline
Star & $Q$ & $p$-mode & $p$-mode & $p$-mode & Frequency & Amplitude\\
& (day) & [22] & [23] & [24] & (cycles/day) & (mag.)\\
\hline
2MASS 00044615+4936439 & 0.024(1) & 1 & 1 & 1 & 8.443 & 0.026\\
2MASS 10215638-3326137 & 0.021(1) & 1 & 1 & 1 & 11.798 & 0.014\\
2MASS 15515693-7759002 & 0.037(1) & F & F & F & 14.589 & 0.036\\
2MASS 07513202+0526526 & 0.035(1) & F & F & F & 15.544 & 0.026\\
\hline
\end{tabular}
\end{center}
\label{Q-F-Amp}
\end{table*}

%%%%%%%%%%%%%%%%%%%%%%%%%%%%%%%%%%%%%%%%%%%%%%%%%%
\vspace{0.6cm}
\section*{Data Availability}
The extracted times of maxima from the TESS space-based data will be made available on request.

%%%%%%%%%%%%%%%%%%%%%%%%%%%%%%%%%%%%%%%%%%%%%%%%%%
\vspace{0.6cm}
\section*{Acknowledgments}
We have made use of data from the European Space Agency (ESA) mission Gaia, processed by the Gaia Data Processing and Analysis Consortium (DPAC, \url{https://www.cosmos.esa.int/web/gaia/dpac/consortium}). This work includes data from the TESS mission observations. The NASA Explorer Program provides funding for the TESS mission.
This study uses information from the ASAS-SN variable stars database.

%%%%%%%%%%%%%%%%%%%%%%%%%%%%%%%%%%%%%%%%%%%%%%%%%%
\vspace{0.6cm}
\section*{ORCID iDs}
\noindent Atila Poro: 0000-0002-0196-9732\\
Hossein Azarara: 0009-0003-2631-6329\\
Ahmad Sarostad: 0000-0001-6485-8696\\
Nazanin Kahali Poor: 0009-0007-5785-7303\\
Razieh Aliakbari: 0009-0007-8508-2357\\
Sadegh Nasirian: 0009-0001-8140-1505\\
Sahar Momeni: 0009-0005-3133-9079\\

%%%%%%%%%%%%%%%%%%%%%%%%%%%%%%%%%%%%%%%%%%%%%%%%%%%%

\end{document}